\documentstyle [12pt,aasms4]{article}
\journalid{}{}
\articleid{}{}

\begin{document}
\def\lax    {\ifmmode{_<\atop^{\sim}}\else{${_<\atop^{\sim}}$}\fi}
\def\gax    {\ifmmode{_>\atop^{\sim}}\else{${_>\atop^{\sim}}$}\fi}
\def\gtorder{\mathrel{\raise.3ex\hbox{$>$}\mkern-14mu
             \lower0.6ex\hbox{$\sim$}}}
\def\ltorder{\mathrel{\raise.3ex\hbox{$<$}\mkern-14mu
             \lower0.6ex\hbox{$\sim$}}}

\title{X-RAY SPECTRAL FORMATION IN A CONVERGING FLUID FLOW:
 SPHERICAL ACCRETION INTO BLACK HOLES}

\author{
Lev Titarchuk\altaffilmark{1,4}, Apostolos Mastichiadis\altaffilmark{2},
and Nikolaos D. Kylafis\altaffilmark{3}}

\altaffiltext{1}{National Aeronautics and Space Administration, 
Goddard Space Flight Center
(NASA/GSFC), Greenbelt, MD 20771, USA; E-mail: titarchuk@lheavx.gsfc.nasa.gov}
\altaffiltext{2}{Max-Planck-Institut f\"ur Kernphysik, Postfach 10 39 80,
D-69029 Heidelberg, Germany; E-mail: masti@aposto.mpi-hd.mpg.de}
\altaffiltext{3}{University of Crete and Foundation for Research and Technology
- Hellas; E-mail: kylafis@physics.uch.gr}
\altaffiltext{4}{George Mason University/Institute for 
Computational Sciences and Informatics}

\rm

\vspace{0.1in}

\begin{abstract}
We study Compton upscattering of low-frequency photons in a converging flow
of thermal plasma.  The photons escape diffusively and electron scattering
is the dominant source of opacity.  We solve  numerically and approximately 
analytically the equation of radiative transfer in the case of spherical, 
steady state accretion into black holes. 
Unlike previous work on this subject, we consider the inner boundary at a
{\it finite} radius and this has a significant effect on the emergent spectrum.
It is shown that the bulk motion of the converging flow 
is more efficient in upscattering photons than thermal Comptonization,
provided that the electron temperature in the flow is of order a few keV or 
less.  In this case, the spectrum observed at infinity
consists of a soft component coming from those input photons which escaped
after a few scatterings without any significant energy change 
and of a power law which extends to high energies and is made of those 
photons which underwent significant upscattering.  The luminosity of the power
law is relatively small compared to that of the soft component.
The more reflective the inner boundary is, the flatter the power-law spectrum
becomes.   The spectral energy power-law index for black-hole 
accretion is always higher than 1 and it is approximately 1.5 for high
accretion rates.  This result tempts us to say that bulk motion 
Comptonization might be the mechanism behind the power-law spectra seen in 
black-hole X-ray sources.

\end{abstract}


\section{INTRODUCTION}

The problem of Compton upscattering of low-frequency photons in an optically
thick, converging flow has been studied by several researchers.  
Blandford and Payne were the first to address this problem in a series of
three papers (Blandford \& Payne 1981a,b, hereafter BP81a and BP81b
respectively; Payne \& Blandford 1981, hereafter PB81).

In the first paper of the series they derived the Fokker-Planck
radiative transfer equation which took into account
photon diffusion in space and energy, while in  
the second paper they studied the acceleration of photons in a 
radiation-dominated, plane-parallel shock by using the
Fokker-Planck formalism developed in  the first paper. 
In the third paper they solved the Fokker-Planck radiative transfer 
equation in the case of steady state, spherically symmetric,
super-critical accretion into a central black hole with the assumption
of a power-law flow velocity $v(r)\propto r^{-\beta}$ and neglecting 
thermal Comptonization.  For the inner boundary condition they assumed 
adiabatic compression of photons as $r \to 0$.  Thus, their flow extended from
$r=0$ to infinity.  They showed that all emergent spectra have a high-energy, 
power-law tail with index $\alpha=3/(2-\beta)$ (for free fall $\beta=1/2$ 
and $\alpha=2$), which is independent of the low-frequency
source distribution.  

In a extension of the work of BP81b, Lyubarskij \& Sunyaev 
(1982, hereafter LS82) studied Comptonization in a radiation-dominated shock, 
taking into account not only the bulk motion of the electrons but also
their thermal motion. Also
Colpi (1988, hereafter C88) generalized  the work of PB81 by including the 
thermal motion of the electrons along with their bulk motion.  
However, as in PB81, the inner boundary condition was taken at $r=0$ and 
thus neglected the black-hole horizon.
 
The LS82 study was further extended numerically by Riffert (1988, 
see also the related work by Bekker 1988),
who found a self-consistent solution for the photon spectrum, the velocity
and the temperature profile in  the infinite shock.
The spectra derived there exhibit all the range of 
features from the thermal Comptonization dominated case to the bulk motion 
dominated one.  
Mastichiadis \& Kylafis (1992, hereafter MK92) studied the upscattering of low 
frequency photons in an optically thick, spherical accretion
onto a neutron star due to bulk motion only.   The effects of the radiation 
force on the flow and thermal Comptonization were neglected, while the 
neutron-star surface was considered to be entirely reflective.
They showed that the high-energy part of the emergent spectrum is
a power-law with energy spectral index essentially zero.
No high-energy cutoff was found because the Compton degradation term was not 
included in the radiative transfer equation.

In the present work we give an accurate numerical and an approximate 
analytical solution of the problem of spectral formation 
in a converging flow, which takes into account the inner boundary condition,
the free fall motion and the thermal motion of the electrons.
The inner boundary is taken at a {\it finite} radius and the spherical 
surface there is considered to be anything from fully reflective to
fully absorptive. The fully absorptive inner boundary 
mimics a black-hole horizon.
No relativistic effects (special or general) are taken into account
in this work and thus our spectra, though very suggestive, 
make instructive sense only, without
being directly comparable with observations. 
However, our work derives for the first time the spectral-energy power-law 
index as a function of accretion rate. It demonstrates by using 
numerical and analytical techniques that the extended power laws are 
present in the resulting spectra. Thus, it constitutes a significant
step towards an exact treatment of Comptonization in matter accreting 
onto compact objects.
Some of the results presented here have already appeared
elsewhere (Titarchuk, Mastichiadis \& Kylafis 1996).        

In \S~2 we give a qualitative discussion of the problem to be solved
and of the solution that we have obtained.  
In \S~3 we solve numerically and approximately analytically
the radiative transfer equation and discuss the properties of the
emergent spectrum.  
Finally,  we summarize our work and draw conclusions in \S~4.

\section{ QUALITATIVE DISCUSSION}

\subsection[QUALITATIVE DISCUSSION]{ Accretion Flow}

Let $\dot M$ be the rate at which matter is accreting and let its
radial inward speed be 
$$
v_b(r) = c \left[ { {r_s} \over r} \right]^{1/2} ~,
\eqno(1)
$$
where $c$ is the speed of light and $r_s$ is the Scwarzschild radius.
This free-fall velocity profile makes the tacit assumption that the radiation
force on the accreting matter is insignificant, or equivalently, the 
escaping luminosity is much less than the Eddington value.
The Thomson optical depth of the flow from some
radius $r$ to infinity is given by
$$
\tau_T(r)=\int_r^\infty drn_e(r)\sigma_T= \dot m
\left[ { {r_s} \over {r}}
\right]^{1/2} ~,
\eqno(2)
$$
where $n_e(r)$ is the electron number density,
$\sigma_T$ is the Thomson cross section, $\dot m=\dot M /\dot M_E$
and $\dot M_E$ is the Eddington accretion rate defined by
$$
\dot M_E \equiv {L_E\over c^2}={4\pi GMm_p\over \sigma_Tc} ~,
\eqno(3)
$$
where $L_E$ is the Eddington luminosity, 
$M$ is the mass of the central object, $m_p$ is the proton mass
and $G$ is the gravitational constant.
We remark here that $\dot M$ can be significantly larger than $\dot M_E$,
while the escaping luminosity $L$ remains much less than $L_E$, because the
efficiency of converting accretion energy into luminosity is significantly 
less than unity for radially accreting black holes.  Thus, our assumption
in equation (1) of a free-fall velocity profile is justified
(see also \S\ 3.6 below).

For super-critical accretion into black holes ($\dot m\ge1$), it is evident 
from equation (2) that
there should be regions in the flow from which the photons escape
diffusively. Of {\it qualitative} importance in our discussion is the
trapping radius $r_{tr}$ (Rees 1978; Begelman 1979) defined 
by the relation
$$
3 {v_b(r_{tr})\over c}\tau_T(r_{tr})\equiv 1 ~.
\eqno(4)
$$
No {\it quantitative} importance should be given to the trapping radius, because
the mean number of scatterings that photons undergo depends on the photon source
distribution.  

For our problem it is convenient to use not 
the Thomson optical depth $\tau_T$ but the effective optical depth 
$$
\tau^\prime
\equiv 3{v_b(r)\over c}\tau_T(r)
= 3 \dot m { {r_s} \over r} ~,
\eqno(5)
$$
defined such that $\tau^\prime (r_{tr})=1$.
The variable $\tau\equiv\tau^\prime/2$ will replace the radial coordinate 
$r$ in the radiative transfer equation in \S\  3.

\subsection[QUALITATIVE DISCUSSION]{ Photon Energy Gain and Loss}

The physical picture here is the following:  Consider a low energy photon
that finds itself in the accretion flow.  As the photon diffuses outward,
it scatters off the inflowing electrons.  Since this, in most cases, is an
almost head-on collision between a fast moving electron and a low energy 
photon, the photon gains energy on average.
The problem to be solved can then be described qualitatively as follows:
Consider a source of photons with frequency $\nu_0$, or a distribution
around this frequency, such that $h\nu_0 \ll m_e c^2$.  The source
of photons is placed anywhere in the accretion flow between the inner and
outer boundaries of the flow and with any spatial distribution.
For large optical depths in the accretion flow, the photons diffuse in the
flow and end up either in the hole or at infinity.  For small optical
depths (say of order unity), the majority of the input photons escape  
after a few scatterings without 
changing their energies significantly (almost coherently scattered).
However, a small fraction of the input photons undergo 
effective scatterings with 
significant energy change (Hua \& Titarchuk 1995).  
In all cases, the diffusing photons gain energy from the bulk and the thermal 
motion of the electrons.  The objective is to determine the emergent spectrum.

Since the bulk motion involves high speeds, the thermal motion of the electrons
has little effect on the emergent spectrum if $kT_e \ll m_ec^2$ (see below 
for a more quantitative statement).  In such
a case, the asymptotic behavior of the emergent spectrum is a power-law.
The exact value of the power-law exponent depends on the condition at the
inner boundary and the accretion rate.

If the inner boundary is fully reflective (MK92), all input photons escape.
Furthermore, the repeated reflection of the photons at the inner boundary
causes them to scatter many times with the inflowing electrons and the
emergent power law is as flat as it can be.  For very optically thick flows, 
the spectral energy power-law exponent  is almost zero.  

In the opposite case where the inner boundary is fully 
absorptive, as is the case of a black-hole 
horizon, the emergent spectrum consists only 
of the photons that {\it did not} reach the inner boundary.  These photons have 
had fewer chances to collide with the inflowing electrons and thus the emergent 
power law is steeper than in the fully reflective case.  {\it The power-law
exponent is different than that found by previous workers and this is
entirely due to the finite radius of the inner boundary.}

In all cases, the power-law spectrum produced by the bulk motion of the 
electrons has a cutoff at high energies due to Compton recoil.  Below
we investigate the relative importance of the bulk (converging
flow) and the thermal motion of the electrons to the mean photon energy
change per scattering.

There are two contributions to the mean energy gain per scattering of a photon.
The first contribution  is $<\Delta E^{(1)}_{b}>$,
which is  proportional to the first power of $v_b/c$ (BP81a),
and is given by (see Appendix D)
$$ 
<\Delta E^{(1)}_{b}>\approx E {4 \over {\dot m}} ~.
\eqno{(6)}
$$
It is worth noting that an additional factor $1/3$ is present in equation (6)
in the case of plane geometry considered by LS82 and Riffert (1988).
 
The second contribution is proportional to the second power of the electron 
velocity, i.e., $ u^2 = v_{th}^2+ v_b^2$,
where $v_{th}$ and $v_b$ are the thermal and bulk velocity respectively,
and it is given by (see Appendix D and also eq. [15] of BP81a)
$$ 
<\Delta E_{th,b}>=<\Delta E_{th}>+<\Delta E^{(2)}_{b}> 
\approx E{{4(kT_e+m_ev_b^2/3)}\over{m_ec^2}} ~.
\eqno{(7a)}
$$
Thus we have that
$$
{{<\Delta E_{th}>} \over {<\Delta E^{(1)}_{b}>+<\Delta E^{(2)}_{b}>}}
\approx {{4kT_e} \over {m_ec^2\{4\dot m^{-1}+
4[\tau_T(r)/\dot m]^2/3\}}} < { 1\over {\delta}} ~,
\eqno{(7b)}
$$
where 
$$\delta \equiv 1/\dot m\Theta 
=51.1\times T_{10}^{-1}\dot m^{-1} ~, 
\eqno{(8)}
$$
where $\Theta = kT_e/m_ec^2$ and $T_{10}\equiv kT_e/(10 $ keV).  
Hence, the bulk motion Comptonization dominates the thermal one
if  $\dot m T_{10} < 51$. 

The high-energy cutoff of the spectrum can
be understood in the following way: As it is shown in Appendix D, the 
fractional increase in energy of a low-energy photon in its collision with 
accreting electrons of bulk velocity $v_b$ and temperature $T_e$ is given by
$$
{{<\Delta E>_{incr} }\over{ E}}
 \approx 4\dot m^{-1}+ 
(4/3)(\tau_T(r)/\dot m)^2 + {{4 k T_e} \over {m_e c^2}} ~.
\eqno{(9)}
$$
At the same time, the recoil effects cause a fractional decrease in the energy 
of the photon given by (see Appendix D)
$$
{{<\Delta E>_{decr} }\over {E}} \approx - {{E}\over{m_ec^2}} ~.
\eqno{(10)}
$$
When $E \ll m_ec^2$, the recoil effect is negligible, resulting  in a pure 
power-law spectrum.
At high energies and $\dot m >1$
the two effects are comparable and the cutoff in the
spectrum occurs at 
$$
{ {E_c} \over {m_ec^2} } \approx 
4 \dot m^{-1}+ 
(4/3)(\tau_T/\dot m)^2 + {{4kT_e} \over {m_e c^2}} ~,
\eqno{(11)}
$$
or $E_c \approx m_ec^2 [4/{\dot m} +(4/3)(\tau_T/\dot m)^2]$ 
when $kT_e \ll m_ec^2$ and $\tau_T$ is an effective Thomson optical depth
expected to be less than $\dot m$.
Because of the high efficiency of the 
recoil effect at energies of order $m_ec^2$  such photons lose 
their energy completely after a few scatterings (much less than average). 

\subsection[QUALITATIVE DISCUSSION]{Emergent Spectrum}

In problems such as ours, the radiative transfer equation can be solved 
by the method of separation of variables (PB81; C88).
Thus, the problem is reduced to finding the eigenvalue and the eigenfunction 
of the $k$-component of the solution.  
In other words, the solution for the occupation number 
$n(\tau,x)$ is a series of the form
$$ 
n(\tau,x)= \sum_{k=1}^{\infty}c_kR_k(\tau)N_k(x) ~,
\eqno{(12)}
$$    
where $\tau$ and $x$ are the space and energy variables respectively,
$R_k(\tau)$ is the eigenfunction of the space operator, 
$N_k(x)$ is the spectrum of the $k$-mode of the solution
and $c_k$
are the expansion coefficients of the space source distribution over the 
set of the space eigenfunctions $\{R_k(\tau)\}$.
For the $k$-component of the solution [i.e., $c_kR_k(\tau)N_k(x)$],
the spectrum $N_k(x)$ {\it is the same at any place in the converging flow.}
We use  this modification of the method of separation of variables to
find an approximate analytic solution (see \S 3.3 and Appendix E).

 We have solved the radiative transfer equation numerically
because  an exact solution of the 
form (12) cannot be found for all energies and  
because the inner boundary condition depends on 
energy and also the energy part of the radiative transfer 
equation (see eq. [14] below) depends on the space variable (optical depth). 
The numerical solution has shown that the emergent spectrum can be well
approximated by an analytic spectrum consisting of two parts.
The first is made by 
those input photons which escape scattered but with negligible change 
of their energies, while the second is a power
law extending to high energies (see eqns. [21] and [29] below).  
\section{RADIATIVE TRANSFER}
\subsection[RADIATIVE TRANSFER]{The Radiative Transfer Equation and 
its Solution}

Now we proceed to write down and solve the radiative transfer equation.
Consider spherically symmetric accretion onto a compact
object with rate $\dot m$.  The inflow bulk velocity of the electrons is
${\bf v_b}$ and their temperature is $T_e$.  Following BP81a 
(their eq. [15]), 
where the bulk motion  term $v_b^2$ is not neglected with respect to the 
thermal one as it is in their eq. [18]), we write for the 
occupation number $n(r, \nu)$
$$
{ {\partial n} \over {\partial t} } +
{\bf v_b} \cdot \nabla n =
\nabla \cdot \left[ { c \over {3 \kappa} } \nabla n \right] +
{1 \over 3} \nu { {\partial n} \over {\partial \nu} } \nabla \cdot {\bf v_b} 
~~~~~~~~~~~~~~~~~~~~
$$
$$
~~~~~~~~~~~~~~~~~~~~
+ { 1 \over {\nu^2} } { {\partial} \over {\partial \nu} }
\left[ { {\kappa h} \over {m_e c} } \nu^4
\left( n + { {kT_e+m_ev^2_b/3} \over h} { {\partial n} \over {\partial \nu} }
\right) \right] + j(r, \nu) ~,
\eqno(13)
$$
where $h$ is Planck's constant, $\kappa(r) \equiv n_e(r)\sigma_T$
is the inverse of the scattering mean free path, 
and $j(r, \nu)$ is the emissivity.
Substituting for the inflow velocity ${\bf v_b} = -v_b(r) {\bf \hat r}$, where
${\bf \hat r }$ is the radial unit vector, and taking into account the
spherical symmetry, equation (13) becomes in steady state
$$
\tau { {\partial^2 n} \over {\partial \tau^2} }
- \left( \tau + {3 \over 2} \right) 
{ {\partial n} \over {\partial \tau} } = 
{1 \over 2} x { {\partial n} \over {\partial x} }
- { 1 \over {2\delta_b} } { 1 \over {x^2} } 
{ \partial \over {\partial x} }
\left[ x^4 \left( f_b^{-1}n + { {\partial n} \over {\partial x} } 
\right) \right]
- {{\dot m} \over {2}}  
{ j \over {\kappa c} } ~,
\eqno(14)
$$
where $x\equiv h\nu/kT_e$, $\tau\equiv\tau^\prime/2$ defined by equation (5)
and 
$$
\delta_b^{-1} \equiv \delta^{-1}f_b(\tau) ~,
\eqno(15)
$$
where $f_b(\tau)=1+(v_b/c)^2/(3\Theta)$.

The  dimensionless spectral energy flux $F(r, \nu)$ (PB81, C88) is written 
in the new variables as 
$$
F(\tau,x) = x^3 \left[
{{2\tau} \over {3\dot m} }\right]^{1/2}
\left({{\partial n}\over{\partial \tau}}
+{1\over3}x{{\partial n}\over{\partial x}}\right) .
\eqno(16)
$$
Note that the normalization factor
$F_0=2\pi \times 10^{23} (kT_e/{\rm keV})^3 {\rm erg}~{\rm cm}^{-2}{\rm 
s}^{-1}{\rm keV}^{-1}$ should be introduced in order to express 
$F(\tau,x)$ in flux units.

There are two boundary conditions which our solution must satisfy.
The first is that the total spectral flux integrated over the outer boundary
should depend only on $x$ as $\tau \to 0$. Thus,
$$ 
F(\tau_o, x)= C(x)\tau^2_o ~~~{\rm as}~~~~
\tau_o \to 0 ~,
\eqno(17)
$$
where the subscript $o$ means outer boundary.
The second boundary condition is that we have a boundary 
with albedo $A$ at some radius $r_b$, or equivalently at an
effective optical depth $\tau_b$, defined by
$$
\tau_b \equiv { {\tau_b^\prime} \over 2} = {3 \over 2} { {v_b(r_b)} \over c}
\tau_T(r_b) ~.
\eqno(18)
$$
The reason for considering a general albedo $A$ and not 
only the case $A=0$ (appropriate 
for black holes) is to be able to compare our results with those obtained
in previous work.
The net energy flux through this surface is
$$
F(\tau_{b},x) =
-x^3\left({{1-A}\over{1+A}}\right){n\over2} ~.
\eqno(19)
$$  
>From the setup of the problem, $\tau_b$ is the largest value that the variable
$\tau$ can obtain. 
\par
In a recent preprint, Psaltis and Lamb (1996) rederived the photon kinetic
equation for Comptonization, correcting several inaccuracies in the literature.
Expansion of the photon kinetic equation in the subrelativistic regime
$v_b/c<1$ over the spherical harmonics leads to equation (14) 
(or eq. [15] of BP81a) for the zeroth moment of the occupation number.  

\subsection[RADIATIVE TRANSFER]{Numerical Solution}

Equation (14), satisfying the boundary conditions (17) and (19), has been 
solved numerically and the emergent spectrum has been computed.  The method 
that we used is the relaxation method for the solution of the boundary 
problem for elliptic partial differential equations (e.g., Press et al. 
1992). The solution of the stationary problem is searched as 
the equilibrium solution  of the approriate initial value problem. 

An interesting example of a space source distribution,
mimicking the external illumination of a converging flow by the
low-energy radiation of an accretion disk, is given by
$$
S(\tau) = S_0 \tau^2\exp(-\tau_T(\tau)/\mu) ~,
\eqno(20)
$$
where $S_0$ is a normalization constant.  An external illumination produces an 
exponential source distribution over the converging flow. 
The factor $\tau^2\propto r^{-2}$ takes into 
account the space dilution of the radiation, while $\mu$ is the cosine
of the angle of incidence at $r = r_s$.
The physical meaning of this space 
source distribution is the following:  Consider a source of soft photons
{\it outside} the converging flow and at a distance $R_0\gg r_s$ 
from the center of 
the black hole.  The source illuminates from the outside the accretion flow
and the radiation penetrates with an exponential probability to
radius $r$, where it is isotropized by scattering.  Thus, the scattered 
photons at radius $r$ constitute the source of soft photons that get
Comptonized in the accretion flow.  We note here that this does not violate
the condition of no incoming radiation at the outer boundary. This is 
a standard procedure for the reduction of  the illumination
problem to the problem of source distribution within the atmosphere 
(e.g., Sobolev 1975). 

\placefigure{fig1}

A specific example of the emerging
spectra, related to the source space distribution (20) 
and to the source spectral distribution   as a blackbody 
spectrum,  is  shown
in Figure~1. Two of the emergent spectra are presented there. 
The first one (solid line) is a solution of 
equation (14) with the appropriate boundary conditions (17) and (19),
and $\delta_b$ as given by equation (15).  
The second spectrum (dashed line) is for the case 
where the second order effect of the bulk motion has been neglected
in the radiative transfer equation. 
Neglecting the  $(v_b/c)^2$ term  in equation (14) transforms
it into  equation (18) of BP81a. 
It is evident from Figure 1 that the emergent spectral energy flux
consists essentially of two components: The first one is 
due to the majority of the input
photons which escape without significant 
energy change, while
the second one is the flux of those relatively few photons 
which undergo effective repeated scatterings and produce 
an extended power law. 
At the same time, the inclusion of the second order bulk effect 
produces an identical shape of the hard tail (for energies higher than
10 keV) but with a larger photon contribution 
which, in addition, 
is extended to higher energies.
This is because the second order effect of the bulk motion
$(v_b/c)^2$ increases the effectiveness of photon upscattering.

\subsection[RADIATIVE TRANSFER]{Analytic Approximation of the Solution}

In view of the power-law that is evident in the emergent spectrum in Figure
1, it is natural to seek an approximate analytic solution of equation
(14) where  a power law is a part of the solution.  
First, a ``solution'' of the form (12) (so called Green's function)
for equation (14)  
with the boundary conditions (17) and (19)  
is derived (for a justification see \S\ 3.4 below).
Three assumptions are made in order to separate
the energy and space variables (i.e., to present the solution as series [12]).

The first one is that we neglect the bulk motion term $(v_b/c)^2$ 
in equation (14) which
is a function of the space variable $\tau$. It is shown in Figure 1 
that the inclusion  of this term  in the equation
does not change the  spectral shape of the high-energy part of the spectrum.

The second one is that  we assume that the Green's function 
(i.e., the spectrum for $\delta-$function energy injection) 
is a power law up to a high energy cutoff. 
In Appendix B the energy Green's function is found (eq. [B1]) and
the validity of the power-law assumption is
demonstrated in Appendix C  (see also Fig. 4a, curve 1).  

The third one is that, 
in order to get the complete system of eigenfunctions, 
we keep the same boundary condition (eq. [A4] of Appendix A) for all of them. 

In Appendix  A the space eigenvalue problem is solved 
under these assumptions
and the transcendental 
equation (A6)  is derived for the spectral index determination.
The approximate analytic solution 
for the spectral flux can be written as
the series (see eqns. [12] and [16]) 
$$
F^{tot}(0,x)=4\pi r^2F(0,x)
= C_{F}\sum_{k=1}^{\infty}{{c_k}\over{2\lambda^2_k-3}} I_k(x,x_0) ~,
\eqno(21)
$$ 
where
$$
C_{F}= 10\pi r_s^2(3/2)^{3/2} \dot m^{5/2} ~,
$$
$c_k$ are the expansion coefficients of the space source distribution
$S(\tau)$ given by equation (B9),
$\lambda_k^2$ are the eigenvalues 
of the boundary value problem (A4-A8), while $I_k(x,x_0)=x^3G_k(x,x_0)$
is the energy Green's function (with $G_k(x,x_0)$ the occupation number
Green's function of the  Comptonization equation [B1]).
In the case where $x_0\ll 1$ (i.e., the input photon energy
$E_0\ll kT_e$), $I_k$ is given by
$$
I_k=\displaystyle{{b_k}\over {2\mu_k x_0}}
 \cases{\displaystyle{\left({x \over {x_0}}\right)^{\alpha_k+3+\delta}},
~~&{{\rm for}~~$x\leq x_0 ;$} \cr
\displaystyle{{e^{-x}}\over{\Gamma(2\mu_k)}}
\displaystyle{\left({x \over x_0}\right)^{-\alpha_k}}
\displaystyle{\int_0^\infty e^{-t}(x+t)^{\alpha_k+3+\delta}
t^{\alpha_k-1}dt},~~ &{{\rm for}~~$x\geq x_0 .$}}
\eqno(22)$$
Here
$ b_k=\alpha_k(\alpha_k+3+\delta)$, 
$\mu_k= {1\over2}[(\delta-3)^2 + 8\lambda_k^2\delta]^{1/2}$, 
and the spectral index is determined by
$$
\alpha_k= {1\over2} [(\delta-3)^2 + 8\lambda_k^2 \delta]^{1/2}-
(3+\delta)/2 ~, 
\eqno(23)
$$ 
The normalization of $I_k(x,x_0)$ is chosen in such a way as to keep the
photon number equal to $1/x_0$. 

The above relations correspond to a monochromatic photon injection
with energy $E_0$.
In order to get the resulting spectrum for an arbitrary source spectrum
$g(E_0)$ one has to convolve this with the energy Green's function.  Thus,
$$
J_k(E)=\int_0^{\infty}I_k(E,E_0)g(E_0)dE_0 ~.
\eqno(24)
$$

\placefigure{fig2}

We have checked the numerical versus the approximate analytic solution
for a wide range of $\dot m$ and blackbody spectra
of temperature $T_{bb}$.
In Figure 2 we show a specific example of 
the emergent spectral energy flux for $\dot m =3$, $kT_e = 1$ keV, 
$kT_{bb} = 1$ keV and $r_b = r_s$.   
The solid curve is the emergent spectrum found from the
numerical solution of equation (14) without the $v_b^2$ term 
and the dashed curve is our approximate analytic solution.  
As in Figure 1, here
also the emergent spectrum consists of two components:  a) The bulk of 
the soft input photons that escaped from the accretion flow without any 
significant change 
of their energies.  b) A relatively small fraction of the soft photons that is
comptonized to high energies and thus forming  the extended power law.
It is worth noting the discrepancy between the two solutions at 
the high energy cutoff. Our assumption regarding a power law shape
of the Green's function is certainly violated around the high energy
cutoff $E_c$, where the recoil effect becomes important.

\subsection[RADIATIVE TRANSFER]{Properties of Emergent Spectra}

The upscattered
power-law spectrum with the high energy cutoff is accurately described 
by the first term of series (21).
The contribution of the different modes is clearly seen in Figure 3.
The high energy part of the spectrum is
determined by the first mode only, but the low-energy bump is formed
as a result of the summation of all modes.

\placefigure{fig3}

In a realistic situation where radial and disk accretion into a black hole
occur simultaneously,
expression (20) is only a very rough approximation, but this introduces no
qualitative differences in the emergent spectra.
A space source distribution of soft input photons different than that in
equation (20) affects only the bump of soft photons seen  and
leaves unaffected the power law.  The deeper the soft photons are injected,
the smaller the bump. The shallower
the injection is, the fewer the photons that get upscattered into the
power law (Fig. 4a-b).

\placefigure{fig4}

In Figures 4a and 4b the analytic emergent spectra are presented 
for  different space source distributions.  In Figure 4a,
curve 1 is for a distribution proportional
to the first eigenfunction (eq. [A3], with $\hat C=0$).
Curves 2 and 3 are for a distribution appropriate for external illumination 
(eq. [20]) with illumination cosines $\mu=1$ and $\mu=0.3$, respectively.  
It is evident that the spectral slope of the high-energy power law 
is independent of the space distribution of low-energy photons
and, as we shall show in \S\ 3.5, it is determined by the mass accretion rate 
$\dot m$ only. In addition we want to draw the attention of the reader 
to curve 1 in Fig. 4a. This curve represents the Green's function (22).
It is clearly seen that a pure power law describes the spectrum 
perfectly over six energy decades. It can also be proved rigorously that,
despite the presence of the exponential in front of the integral 
in equation (22), the power law can 
extend to $x\gg 1$. As it is shown in Appendix C,
the integral in equation (22) is proportional
to $e^x$ for $x\ltorder \delta$, while 
it is proportional to $x^{\alpha_1+3+\delta}$ for 
$x \gtorder \delta$. Thus, for $kT_e \ll m_ec^2$,
the power law extends to energies 
of order $m_ec^2$ independently of the electron temperature. 
For very high energies, $x \gtorder \delta$, 
the spectrum exhibits an exponential cutoff.

Now we proceed   
with the mathematical analysis of the properties 
of the emergent spectra.  
The source term $j(\tau, x)$ in the right hand side of equation (14) 
can be represented in the factorized form  $j(\tau, x)= S(\tau)g(x)$
without any loss of generality.  For example,
$S(\tau)$ can be given by equation (20) and $g(x)$ can be  a blackbody 
spectrum.  
If one neglects for the moment the energy change in the process of scattering
(i.e., assuming coherent scattering), then   
the radiative transfer equation can be written in operator
form as 
$$ L_{\tau}n_{coh} = -{{\dot m}\over {2}} [g(x)/x^3]S(\tau) ~,
\eqno(25)
$$
where 
$$L_{\tau}=\tau { {d^2 } \over {d \tau^2} }
- \left( \tau + {3 \over 2} \right) 
{ {d } \over {d \tau} }
\eqno(26)$$
is the space operator (see eq. [14]).   
The solution of this equation with the appropriate
boundary condition (A4) is given by the series 
$$
n_{coh}(\tau)=
{ {\dot m} \over {2} }
{ {g(x)} \over {x^3} }
\sum_{k=1}^{\infty}{{c_k}\over{\lambda_k^2}}R_k(\tau) ~.
\eqno(27)
$$
Thus, the flux reads [cf eq. [21])
$$
F^{tot}_{coh}(0,x)=
{ {C_F} \over 2}
\sum_{k=1}^{\infty}{{c_k}\over{\lambda^2_k}} g(x) ~.
\eqno(28)
$$

In the case of bulk motion dominated Comptonization, $\lambda_k^2\gg 3/2$ 
for all $k \geq 2$ (see, e.g., Fig. 5),
and as a result of this 
the corresponding energy Green functions $I_k(x,x_0)$
are steep and act approximately as $\delta-$functions on the input
spectrum.  
Thus  we can approximate the emergent
upscattered spectrum as the following combination
$$
F^{tot}(0,x) \approx {{C_{F}}\over{2}}c_1
\left[{{J_1(x)}\over{\lambda_1^2-3/2}}-
{{g(x)}\over{\lambda_1^2}}\right] +F_{coh}^{tot}(0,x) ~,
\eqno(29)
$$
where $J_1(x)$ is given by the convolution shown in equation (24).
Equation (29) means that the emergent
spectrum can be written approximately as the sum of two components.  The
first component is the difference between the first terms of the full and the
coherent solutions (i.e., it represents the upscattered component), 
while the second component is the fully coherent solution.

For the input distribution (20), it is straightforward to derive that the 
emergent spectral flux is
$$
F^{tot}(0,x) \approx 
F_0 \left\{ c_1\left[{{J_1(x)}\over{\lambda_1^2-3/2}}-
{{g(x)}\over{\lambda_1^2}}\right] +c_{coh}g(x)\right\} ~,
\eqno(30)
$$  
where $F_0$ is a normalization factor in flux units,
$$
c_{coh}=\sum_{k=1}^{\infty}{{c_k}\over{\lambda^2_k}} 
=(2/5)\pi^{1/2}\exp(\eta^2)~{\rm erfc}(\eta) ~,
\eqno(31)
$$
and 
$$
\eta=[\dot m/6]^{1/2}/\mu ~.
\eqno(32)
$$
 
We note here that the approximation of the spectrum presented in equation 
(29) is valid even if the diffusion approximation is not valid.  The only
requirement is that the contribution of the upscattering in the emergent 
spectrum is  efficient enough to produce the power-law hard tail. 
 In such a case, the first expansion coefficient 
$c_1$ should be found by using the first eigenfunction of the kinetic equation
(and not of the Fokker-Planck equation, as in the case of diffusion), while the
coherent component $F_{coh}^{tot}(0,x)$ is the emergent spectrum when no
energy change occurs during scattering (Titarchuk \& Lyubarskij 1995).

\subsection[RADIATIVE TRANSFER]{Power-law Spectral Index}

In order to obtain the eigenvalues of the space operator,
the transcendental equation (A6) is solved numerically 
and the results are presented in Figure 5.
The transcendental equation (A6) 
is factorized in the limit of $\dot m \gg 1$ (see eq. [A8]) and the
roots are determined from the factors.
The dotted line in Figure 5 is the root of the
first factor of equation (A8).
As it also  can be seen,
the solid line and the dotted line perfectly
match each other  in the wide range of  
$1 < \dot m \ltorder 15$, while the two start deviating for larger
values of $\dot m$. This deviation shows the limits of validity and accuracy
of the factorization.

\placefigure{fig5}

The asymptotic expression ($\dot m \gg 1$) for the root of the first factor
of equation (A8) is
$$
\lambda_*^2\approx {3\over2}+ {3\over4}{{1-A}\over{1+A}}
\left[ {r_b\over{r_s}} \right]^{1/2} ~.
\eqno(33)
$$ 
In fact this asymptotic expression for $\lambda_*^2$
describes the numerical value $\lambda_1^2$
with an error of less than 10\% (see Fig. 5).

Having obtained the eigenvalues of the space operator, we can
calculate the 
energy spectral index $\alpha_1$ of the dominant term.
As it can be seen from Figure 6a, this quantity demonstrates
a weak dependence on the electron temperature for $kT_e \le 3$ keV. 
It is evident that in the wide range of $\dot m $ ($2 \ltorder \dot m \ltorder
15$) we have $1<\alpha_1< 1.5$, which is different than what previous
workers have found (PB81, C88).  

\placefigure{fig6}

For $kT_e\ll m_ec^2$ such that $\delta\gg 1$,
equation (23) can be expanded to first order in $1/\delta$, giving
$$
\alpha_1 \approx 2\lambda_1^2-3 ~,
\eqno(34)
$$
The accuracy of the above formula is better than 10\%
when the temperature is 1~keV and it becomes better than 1\%
in the case where the temperature is 0.01 keV.

As long as $\delta\gg 1$, bulk Comptonization effects dominate the thermal 
ones and the spectral index $\alpha_1$ of the fundamental mode is 
weakly dependent on the temperature of the electrons.

Combining equations (33) and (34) we find that for $\dot m \gg 1$ and
$\delta \gg 1$ the dominant spectral index becomes
$$
\alpha_1 \approx {3\over2}{{1-A}\over{1+A}}
\left[{r_b\over{r_s}}\right]^{1/2} ~.
\eqno(35)
$$ 
with an error of less than 10\% for $kT_e=1$ keV.

The dependence of the spectral index $\alpha_1$ on the albedo $A$ is 
displayed in Figure 6b.  As we have already pointed out, for
$A=0$ the power-law index is in the range
$1<\alpha_1< 1.5$ for $\dot m \gtorder 2$.  
For the fully reflective inner boundary
case ($A=1$), $\alpha_1 \to 0$ for $\dot m \gg 1$, as it was found
by MK92.  For all values of $A$ we have taken $kT_e=1$ keV and
$r_b=r_s$.

It is worth noting that the first term of equation (21)  
vanishes exponentially when $\tau_b\to \infty$, 
since the expansion coefficient $c_1$ goes 
exponentially to zero. This happens because the normalization $H_1$ 
of the fundamental space function
$\tau^{5/2}\Phi(-\lambda_1^2+5/2,7/2,\tau)$ (see eqns. [A3], [B9],  [B10]) 
grows exponentially when $\tau_b \gg 1$ (Abramowitz \& Stegun 1970). 
{\it Thus, series (21) converges to the solution found by PB81 and C88
when $\tau_b\to \infty$.}

\subsection[RADIATIVE TRANSFER]{Comptonization Enhancement Factor}

A relation can be found between the luminosity of the low-frequency source
and the luminosity emerging from the converging flow.  Let us use
monochromatic, low-frequency, dimensionless, input luminosity
with unity normalization, i.e., 
$L_{0}=\int_0^\infty \delta(x-x_0)dx=1$. The luminosity of any
energy component $I_k(x,x_0)$ (eq. [21]) is given by
$$
L_k=\int_0^\infty I_k(x,x_0)dx ~.
\eqno(36)
$$
Hereafter the index $k$ has been dropped for simplicity.

One can get by using equation (22) and the integration technique (e.g., 
Sunyaev \& Titarchuk 1985), 
$$
L=\displaystyle{{L}\over{L_{0}}}=
\displaystyle{\alpha(\xi-1)}
 \cases{\displaystyle{{{1}\over {\xi(\alpha-1)}}
\left(1- {{\xi}\over{\xi+\alpha-1}}
x_0^{\alpha-1}\right)},
~~&{{\rm for}~~$\alpha \geq 1$;} \cr
\displaystyle{{\Gamma(\xi)\Gamma(\alpha)
\Gamma(1-\alpha)(1-x_0^{1-\alpha})x_0^{\alpha-1}}\over
{\Gamma(\alpha+\xi)}},~~ &{{\rm for}~~$\alpha\leq 1$,}}
\eqno (37)
$$
where $\xi=\alpha+4+\delta$.
This result is a generalization of the result of 
Sunyaev \& Titarchuk (1980) 
for the converging inflow
case and expression (37) produces  a continuous transition through 
the value $\alpha =1$. 

The important conclusion which can be deduced from equation (37) 
is that the low-frequency 
source flux is amplified only by a factor $\sim$ 3 (for $\alpha_1\approx 1.5$)  
due to Comptonization in the accretion flow into a black hole. Thus,
the effectiveness of the bulk motion Comptonization is rather weak.
It is easy to show that the upscattering effect is negligible  
($L/L_{0}\approx 1$) for the $k$ mode with $k\geq 2$,
because the spectral indices $\alpha \geq 2$ (eq. [34] holds 
for all modes when the temperature does not exceed 1 keV). 

In \S 2 we assumed a free-fall velocity profile. 
Since the energy gain 
due to the bulk motion Comptonization is not bigger 
than a factor of 3, it follows 
that we can safely neglect the effects of the radiation force in our 
calculations if the injected photon flux in the converging inflow 
is of order of a few percent of the Eddington luminosity.

\section{DISCUSSION AND CONCLUSIONS}

We have studied Compton upscattering
in the case of thermal plasma infalling radially into a 
compact object.
The flow is considered to be finite (as opposed to an infinite one considered 
in the related work of PB81 and C88) by taking as inner boundary a totally 
absorptive surface to simulate spherical accretion into a black hole.
The emergent spectrum is calculated by 
solving the radiative transfer equation in the diffusion 
approximation with a soft source of input photons. 

We have considered two different expressions of the transfer equation.
In the first case we have included the second order effect of the
bulk motion (eq. [14]), while in the second case we have neglected it
reducing the equation to the one solved in PB81 and C88 --note, however,
the different boundary conditions. We have solved the equation in 
the first case numerically and in the second case both
numerically and approximately analytically. While, formally speaking,
the second order bulk effect should not be neglected,
by comparing our numerical solutions
for the two cases we were able to conclude
that the inclusion of this effect does not 
significantly change the properties of
the emergent spectrum. Thus solving approximately analytically the transfer 
equation when the second order bulk effect has been neglected
enabled us to study in detail the properties of the emergent spectrum.
This spectrum consists of two components, irrespective of the assumptions 
on the second order bulk effect:  The first component consists of those
soft input photons that escaped  without 
any significant energy change, while the second is a power law that extends
to high energies. The power law spectral index is not sensitive 
to whether the second order bulk effect is included in the
transfer equation or not (see Fig.1). Thus, by using the 
analytic solution we were able to determine the spectral index
as a function of mass accretion rate $\dot m$ and we showed
that this lies between 1 and 1.5 for a wide range of $\dot m$;
furthermore we showed that the spectral index is largely independent of the 
electron temperature as long as this is below 1 keV (Fig. 6a). 
Therefore  we have found that {\it the finite extent of the converging 
flow has crucial effects on the emergent spectrum for moderately 
super-Eddington mass accretion rates}.  For extremely 
large mass accretion rates, our solution tends to the solution found by
PB81 and C88.

Our approach cannot determine accurately
the exact position of the high energy cutoff. 
When the second order effect of the bulk motion 
is included, the numerically obtained spectra show that the cutoff is of order 
of the electron rest mass. This is further confirmed by Monte-Carlo
calculations (Kylafis \& Litchfield 1997, in preparation).
If the second order
bulk effect is neglected, then the cutoff is of order of $m_ec^2/\dot m$ as this
is confirmed by numerical and analytical estimates.
At any rate, both formalisms show {\it qualitatively} the same 
result, i.e., that bulk motion Comptonization produces a power law
which extends to very high energies, of order the electron rest mass.

Our spectra hint at what may be happening in candidate black-hole X-ray
sources.  We find bulk motion Comptonization as an exciting possibility for
the formation of extended power laws in the spectra of black-hole X-ray
sources.

\section{ Acknowledgements}

We thank Peter Meszaros for reading and evaluating the present paper.
We also thank an anonymous referee for comments on an earlier version of this
paper.  L.T. would like to acknowledge support from an NRC grant,  NASA grant
NCC5-52 and from Sonderforschungsbereich 328 during his visit to MPIK.
N.K. acknowledges support from EU grant CHRX-CT93-0329.
L.T. and A.M. thank the staff of the Foundation for Research and 
Technology-Hellas for their hospitality.
Also L.T. thanks Sandip Chakrabarti, 
John Contopoulos and Thomas Zannias  and A.M thanks John Kirk for 
extensive discussions. 
This work was partially supported by the Deutsche Forschungsgemeinschaft
under Sonderforschungsbereich 328. 

\newpage

\appendix

\section{THE SPACE BOUNDARY CONDITIONS
AND THE EIGENVALUE AND EIGENFUNCTION PROBLEM} 

In this Appendix we  solve the radiative
transfer equation (14) taking into account the boundary
conditions (17) and (19), and under the assumptions stated in \S\ 3.3.
We are looking for 
a general solution of the form 
$$ 
n(x,\tau)=R(\tau)N(x) ~.
\eqno({\rm A}1)
$$ 
The space part of equation (14) can be written as
$$
\tau{{d^2R}\over {d\tau^2}}-(\tau+3/2){{dR}\over{d\tau}}+\lambda^2 
R=0 ~,
\eqno({\rm A}2)
$$
and its solution is (see, e.g., Abramowitz \& Stegun 1970)  
$$ 
R(\tau)=\hat C\Phi(-\lambda^2,3/2,\tau)+
C\tau^{5/2}\Phi(-\lambda^2+5/2,7/2,\tau) ~,
\eqno({\rm A}3)
$$
where $\Phi(a,b,z)$ is the confluent (or degenerate) 
hypergeometric function and $\hat C$ and $C$ are arbitrary constants.
Here $\lambda^2$ are the eigenvalues of the problem, which we determine 
below. One can show that the boundary condition 
(17) implies that $\hat C=0$; otherwise, as $\tau\to 0$, $F(\tau, x)\propto
\tau^{1/2}$, contrary to equation (17).

In Titarchuk \& Lyubarskij (1995) it is proved that 
a power law is the exact solution of the radiative transfer kinetic
equation up to the exponential turnover where 
downscattering due to the recoil effect becomes important.
Thus we assume that the energy part of the first term is a power law     
$N(x)\propto x^{-\varepsilon_1}$ up to the exponential cutoff.
Note that the occupation number spectral index $\varepsilon_1$ is related to
the index $\alpha_1$ of the spectral energy flux by $\varepsilon_1 =
\alpha_1 +3$ (see, e.g., eq. [19]).  
The boundary condition (19) implies for $R_1(\tau)$ 
at $\tau=\tau_b$ that
$$
{{dR_1}\over{d\tau}}-{{\hat\varepsilon_1}\over3}R_1=0 ~,
\eqno({\rm A}4)
$$
where 
$$
\hat\varepsilon_1= \varepsilon_1- {{3(1-A)} \over {2(1+A)}} 
\left[ {{r_b} \over{r_s}} \right]^{1/2} ~.
\eqno({\rm A}5)
$$

This boundary condition, along with equation (A3), implies 
that the eigenvalues are the roots of the equation
$$ 
\left({5\over2}-{{\hat\varepsilon_1}\over3}\tau_b\right)
\Phi(-\lambda^2+5/2,7/2,\tau_b)
+ {{5-2\lambda^2}\over7}\tau_b\Phi(-\lambda^2+7/2,9/2,\tau_b)=0 ~.
\eqno({\rm A}6)
$$
This can be further written as
$$\Phi(-\lambda^2+5/2,7/2,\tau_b)\left[
 \left({5\over2}-{\hat \varepsilon_1\over3} \tau_b \right)
+{{5-2\lambda^2}\over{7}}\tau_b
{{\Phi(-\lambda^2+7/2,9/2,\tau_b)}\over{
\Phi(-\lambda^2+5/2,7/2,\tau_b)}}\right]=0 ~. 
\eqno({\rm A}7)
$$
Using the asymptotic form of the confluent hypergeometric function
$\Phi(a,b,z)$ for large argument $z\gg 1$ 
(Abramowitz and Stegun 1970, eq. [13.1.4])
$$\Phi(a,b,z)={{\Gamma(b)}\over{\Gamma(a)}}e^{z}z^{a-b}
[1~+~O(z^{-1})] ~,
$$
and representing $\Gamma(a)$ as the product 
$${\Gamma(a)}=\Gamma(a+n)/ \prod_{i=0}^{n-1}(a+i)
~~~~~~~{\rm for}~n=1,2,...,
$$
we obtain the asymptotic form of equation (A7) which is valid for
$\tau_b\gg\lambda^2$
$$
\left({{\hat \varepsilon_1}\over3}-1 +O(\tau_b^{-1})\right)
\prod_{n=2}^{\infty}\left({{2n+1}\over{2}}-\lambda_n^2\right)=0 ~.
\eqno({\rm A}8)
$$ 
>From the first term of the product we get
$\hat\varepsilon_1\approx3$ which implies (eq. [A5])
$\varepsilon_1\approx3+ 1.5[(1-A)/(1+A)](r_b/r_s)^{1/2}$. 
The other roots 5/2, 7/2, ..., (2n+1)/2, ... are  related with
the  second, third and $n$th term of the above product respectively. 

Two more steps are required to calculate the emergent spectrum.  The first one
is the determination of the energy Green's function and this is presented in 
Appendix B.  The second one is the proof that the spectrum is a power law
up to the exponential cutoff; this is done in Appendix C.

\clearpage

\section{THE SOLUTION OF THE ENERGY EQUATION AND THE 
EXPANSION COEFFICIENTS OF THE SPACE SOURCE DISTRIBUTION} 

In this Appendix we obtain the solution of the energy 
part of the radiative transfer equation (14). 
This is derived after substituting
the factorized expression (A1) for the occupation number
into equation (14) and it reads
$$
{1\over{x^2}}{d\over{dx}}
\left[x^4\left({{dN_k}\over{dx}} + N_k\right)\right] 
-\delta x {{dN_k}\over {dx}}= \gamma_k N_k - g(x) ~, 
\eqno({\rm B}1)
$$
where $g(x)$ is the spectral part of source function.
The boundary conditions are
$$ 
x^3N_k\to 0 ~~{\rm when}~~x\to 0~~ {\rm or}~~ x\to \infty ~.
\eqno({\rm B}2)
$$
In order to find the Green's function $G_k(x,x_0)$, the spectral part of 
the source function $g(x)$ is assumed to be proportional 
to a delta function, i.e.,
$$
g(x)=\delta(x-x_0)/x^3 ~,
\eqno({\rm B}3)
$$
and the differential operator of equation (B1) should be rewritten in the 
self-adjoint form $(d/dx)[p(x)(d/dx)]$. The integrating factor
$f(x)=e^{x}x^{2-\delta}$ allows us to get such a form 
[i.e., $p(x)=x^2f(x)$] with $p(x)=e^{x}x^{4-\delta}$.
The  reduction of equation (B1) to the self-adjoint form is followed 
by integration of the left and right side of the equation 
from $x_{0-}$ to $x_{0+}$ and thus the constant $D$ 
of the symmetric form of the Green's function is obtained (see below). 
The Green's function is given by
$$
G_k(x,x_0)=\cases {DN_1(x)N_2(x_0),&if $x\le x_0$;\cr
DN_1(x_0)N_2(x), & if  $x\ge x_0$, \cr}
\eqno({\rm B}4)
$$
where $N_1(x)$ and $N_2(x)$ are the solutions of the homogeneous equation
(B1) [i.e., with $g(x)\equiv 0$]. 
The solutions $N_1(x)$ and $N_2(x)$ satisfy the
boundary conditions (B2) at $x\to0$ and  $x\to\infty$ respectively. 
Hence the constant $D$, expressed through the Wronskian,
$$
W[N_1(x_0),N_2(x_0)]~=~N_1(x_0)N_2^\prime(x_0)-N_1^\prime(x_0)N_2(x_0) ~,
\eqno({\rm B}5)
$$
reads
$$
D~=~-{{f(x_0)}\over{x_0^3p(x_0)W[N_1(x_0),N_2(x_0)]}} ~.
\eqno({\rm B}6)
$$
The solutions   $N_1(x)$
and $N_2(x)$ are expressed in terms of the Whittaker functions   
(Abramowitz \& Stegun 1970) as
$$
N_1(x)=
x^{(\delta-4)/2}e^{-x/2}M_{2+\delta/2,[(\delta-3)^2+4\gamma_k]^{1/2}/2}(x) ~,
\eqno({\rm B}7)
$$
and
$$
N_2(x)=
x^{(\delta-4)/2}e^{-x/2}W_{2+\delta/2,[(\delta-3)^2+4\gamma_k]^{1/2}/2}(x) ~.
\eqno({\rm B}8)
$$
The product of $p(x_0)W[N_1(x_0),N_2(x_0)]$ is independent of $x_0$ and
is equal to $-\Gamma(2\alpha_k+4+\delta)/\Gamma(\alpha_k)$.
In the limit of low energy source photons ($x_0\ll 1$), the Green's function 
$G_k(x,x_0)$ is transformed into equation (22). 

The expansion coefficients  $c_k$ of  
the  space source photon distribution $S(\tau)$,  
where $j / \kappa c = S(\tau)g(x)/x^3 $ (see eq. [14]),
are determined  by 
using the orthogonality of the eigenfunctions (A3). 
Thus we obtain 
$$
c_k=\int_0^{\tau_b} 
[e^{-\tau}\Phi(-\lambda_k^2+5/2,7/2,\tau)S(\tau)]d\tau/H_k(\tau_b) ~,  
\eqno({\rm B}9)
$$
where
$$ 
H_k(\tau_b)=\int_0^{\tau_b} d\tau 
e^{-\tau}\tau^{5/2}\Phi^2(-\lambda_k^2+5/2,7/2,\tau) ~.
\eqno({\rm B}10)
$$

\clearpage

\section{THE ASYMPTOTIC FORM OF THE ENERGY GREEN'S FUNCTION}

In order to study the asymptotic form of the solution (21) when $kT_e \ll
m_ec^2$, the integral of equation (22) should be calculated for indices 
$\omega=\delta+3+\alpha_1\gg 1$ by the steepest descend method.
Introducing a new integration 
variable $t^{\prime}=t\alpha$, the integral 
$$ 
J(x,\alpha,\omega)=\int_0^\infty e^{-t^{\prime}}
(x+t^{\prime})^{\omega}
{t^{\prime}}^{\alpha-1}dt^{\prime} ~,
\eqno({\rm C}1)
$$
is  rewritten in the form
$$
J(x,\alpha,\omega)=\alpha^{\alpha}\int_0^\infty \exp{[-\varphi (t)]}
{{dt}\over {t}} ~.
\eqno({\rm C}2)
$$
Here and below the subscript $1$ is omitted for simplicity. 
The exponent of the integrand in equation (C2), 
$$
\varphi (t)= \alpha[t-{\omega\over\alpha}\ln(x+\alpha t)
-\ln t] ~,
\eqno({\rm C}3)
$$
has a maximum at 
$$
t_0={{\omega+\alpha-x+
[(\omega+\alpha-x)^2+4x\alpha]^{1/2}}\over{2\alpha}} ~.
\eqno({\rm C}4)
$$
The second derivative of $\varphi (t)$ is
$$
\varphi^{\prime \prime}(t)=\alpha 
\left[{{\omega \alpha}\over{(x+\alpha t)^2}} +
{1\over{t^2}}\right] ~.
\eqno({\rm C}5)
$$
Finally the integral $J(x,\alpha,\omega)$ is given 
by the steepest descend formula
$$ 
J(x,\alpha,\omega)=\alpha^{\alpha}{{\exp[-\varphi(t_0)]}\over{t_0}}
\left[{{2\pi}\over{\varphi^{\prime\prime}(t_0)}}\right]^{1/2} ~.
\eqno({\rm C}6)
$$
This formula provides very high accuracy for any index $\alpha >1$,
but for $\alpha\le 1$ the integral 
should be transformed by using integration by parts as follows
$$
J(x,\alpha,\omega)=[J(x,\alpha+1,\omega)-
\omega J(x,\alpha+1,\omega-1)]/\alpha ~.
\eqno({\rm C}7)
$$  
>From the above we find that
$$
J(x,\alpha,\omega)=e^x\Gamma(\omega+\alpha)=e^x\Gamma(2\mu) ~,
~~~~~~~{\rm for}~~x<\omega ~,
\eqno({\rm C}8)
$$  
and
$$
J(x,\alpha,\omega)=\Gamma(\alpha)x^{\omega+3+\alpha} ~,
~~~~~~~{\rm for}~~x>\omega ~. 
\eqno({\rm C}9)
$$  
Thus, the spectral energy
power law $E^{-\alpha}$ extends up to a high energy cutoff.
Beyond this energy, the spectrum falls off exponentially (or has a hump) of the
the form $C_h x^{\delta+3}\exp{(-E/kT_e)}$.

\clearpage
\section{ENERGY GAIN AND LOSS PER SCATTERING}

There is a standard procedure to estimate the energy gain and loss
per scatering (e.g., Prasad et al. 1988).
In order to make these estimates  one has to find 
the response of the energy operator of the kinetic equation
(see eq. [14]) to a delta-function injection of photons. The 
dimensionless photon energy density 
in this case is $x^3n = x \delta(x-x_0)$, where $x=E/kT_e$, $E$ is the 
photon energy, $T_e$ is the temperature of the electrons and $n$ is the
occupation number.
The energy change per scattering $\Delta x$ can then be found by    
multiplification of the kinetic equation by $x^3$ 
and integration over $x$.
Thus we have 
$$
\Delta x=
 \int_0^\infty\left\{ -{ {1} \over {\dot m} }x^4 { {\partial n} \over 
{\partial x} }
+ x \Theta f_b { \partial \over {\partial x} }
\left[ x^4 \left( f_b^{-1}n + 
{ {\partial n} \over {\partial x} } \right) \right]\right\}dx ~,
\eqno({\rm D}1)
$$
where   $f_b=1+(v_b/c)^2/(3\Theta)$,
and $\Theta= kT_e/m_ec^2$.  
Integration by parts gives 
$$
\Delta x=
\left[{ 4 \over {\dot m} } 
+ 4[\Theta+(v_b/c)^2/3]\right]\int_0^\infty x^3n dx-
\Theta \int_0^\infty x^4n dx ~.
\eqno({\rm D}2)
$$
Substituting $x^3n = x \delta(x-x_0)$ in this equation we obtain
$$
\Delta x=\left[{ {4} \over {\dot m} } +
 4[\Theta+(v_b/c)^2/3]- 
 {{E_0}\over{m_ec^2}}\right] x_0 ~, 
\eqno({\rm D}3)
$$
where $E_0$ is the photon energy before scattering and $x_0= E_0/kT_e$.
The first term on the right hand side of equation (D3) is 
the Comptonization to first order in the electron bulk velocity, the second
term is to second order in the electron velocity (thermal and bulk) and
the third term gives the photon energy loss due to the recoil of the electron.
\vfill \eject

\clearpage

\section{SEPARATION OF VARIABLES}

We remind the reader how the method of separation of variables works 
in the case of a general linear operator equation of the form
$$
L_x n+ L_{\tau}n = S(\tau)g(x),~~~~ {\rm for}~~ 0 \le \tau \le \tau_b 
~~{\rm and}~~{0<x<\infty},  
\eqno({\rm E}1)
$$
with boundary conditions $l_on=0$ at $\tau = 0$ and $l_{\tau_b}n=0$ at 
$\tau = \tau_b$, where $\tau$ and $x$ are the space and 
energy variables respectively, and the boundary operators
$l_o$, $l_{\tau_b}$ are  independent of the energy variable (in the case
considered in our present paper, $l_{\tau_b}$ depends on energy).
The eigenfunction $R_k(\tau)$ of the space operator
($L_{\tau}R_k+\lambda_k^2R=0$,  $l_oR_k=l_{\tau_b}R_k=0$)
always exists for the very wide class of operators considered in equation 
(E1).  The eigenfunctions form a complete set, i.e., the space part of the 
source function $S(\tau)$ can be expanded in a series
$S(\tau)=\Sigma c_kR_k(\tau)$, with $c_k$ the expansion 
coefficients.  Having in mind this expansion and also
that $n(\tau,x)$ is looked for as a series 
$$ 
n(\tau,x)= \sum_{k=1}^{\infty} c_kR_k(\tau)N_k(x) ~,
\eqno({\rm E}2)
$$
where $N_k(x)$ is the spectrum of the $k$-mode of the solution, 
one can see that the whole problem is reduced to the solution 
of the equation for the k-energy component $N_k(x)$.  Namely,
$$
\Sigma c_kR_k(\tau) [L_xN_k-\lambda_k^2N_k-g(x)]=0,
\eqno({\rm E}3)
$$     
from which one obtains
$$
L_xN_k-\lambda_k^2N_k-g(x)=0,
\eqno({\rm E}4)
$$
because of completeness of the $\{R_k(\tau)\}$.

In general, when the internal boundary operator
$l_{\tau_b}^x$ depends on energy (as it is in our case, see eqs. [16, 19]), 
the above method of separation of variables cannot be applied.
But in the  particular case of BP81a (their eq. [18]; see also our eq. [14] 
with $\delta_b=\delta$), a power law is  a solution in the range 
of dimensionless energies $x_0<x<\delta$ ($x_0\ll 1$, $1 \ll \delta$). 
This can be checked by simple 
substitution of the product $R(\tau)x^{-\alpha}$ into equation (14)
(where $\delta_b$ has to be replaced by $\delta$)
and into the boundary conditions (eqns. [17] and [19]).
After substitution we get a system of equations for the determination
of the spectral index $\alpha$. In addition, we assume that the photon source
spectrum $g(x)$ has characteristic energies much less than $m_ec^2$
(for example, for a $\delta$-function energy injection, $g(E)=\delta(E-E_0)$
and $E_0\ll m_ec^2$), and therefore there are no source photons with
$E\gg E_0$. Thus, instead of equation (E4), we get in this case 
$$
L_{x}x^{-\alpha}-\lambda^2x^{-\alpha}=(\lambda_{\alpha}^2-\lambda^2)x^{-\alpha}
=0,
$$
or
$$
\lambda_{\alpha}^2-\lambda^2=0,
\eqno({\rm E}5)
$$
for $x>x_0$ and where $\lambda_{\alpha}^2$ is the eigenvalue of the operator
$L_{x}$.  
The internal boundary operator $l_{\tau_b}^x$ is a linear differential operator
with respect to the variables $\tau$ and $x$. The power law $x^{-\alpha}$
is an eigenfunction of this operator, namely
$l_{\tau_b}^x x^{-\alpha}= l_{\tau_b}^{\alpha} x^{-\alpha} $. 
Thus 
$$
l_{\tau_b}^x x^{-\alpha}R(\tau)
= [l_{\tau_b}^{\alpha} R(\tau)]x^{-\alpha}=0
~~~~~{\rm at}~~~~\tau=\tau_b,
$$
or 
$$
l_{\tau_b}^{\alpha}R(\tau)=0
~~~~~{\rm at}~~~~\tau=\tau_b,
\eqno({\rm E}6)
$$
and 
$$
l_oR(\tau)=0 ~~~~~{\rm at}~~~~ \tau=0.
\eqno({\rm E}7)
$$
As it is proved in Appendix C, 
a power law is an eigenfunction of the operator $L_x$
for energies up to a high energy cutoff.
Combining equation $L_{\tau}R+\lambda^2R=0$ and equations (E5)-(E7) we get
a complete system of equations for the determination of the 
spectral index $\alpha$, the eigenvalue 
$\lambda^2$ and the eigenfunction.
Practically, the correctness of 
this approach can be clearly seen from Figure 2, where the numerical 
and the analytical solutions are compared.

It is not our goal to go into  the deep mathematical details 
about the uniqueness of this solution.
We can only remind the reader that there is 
a general theorem for the existence and uniqueness of the solution
for the boundary problem of the partial differential equation
of the second order. This theorem is valid for a wide class of
differental operators $L_x$, $L_{\tau}$, $l_{\tau_b}^x$, $l_o$.

\newpage

\figcaption[fig1.eps]
{Plot of the emergent spectral energy flux versus photon energy for 
$\dot m =3$, $kT_e = 1$ keV, $kT_{bb}=1$ keV, $A=0$, $\mu=0.3$
and the radiative transfer equation is solved numerically. 
The dashed line is for the case where the second order bulk effect
is excluded (see BP81a, their eq. [15]), while the solid line includes
this effect (eq. [14] of the present paper). \label{fig1}}

\figcaption[fig2.eps]
{Plot of the emergent spectral energy flux versus photon energy.
The solid line is identical to the dashed one in  Figure 1, while
the dashed line corresponds to the approximate analytic solution.
\label{fig2}}
 
\figcaption[fig3.eps]
{Plot of the emergent spectral energy flux versus photon energy.
Here $kT_e=2.5$ keV, $A=0$, $T_{e}/T_{bb}=3$, $\dot m=3$.
The emergent spectrum is calculated as a convolution of series (21)
with a blackbody source distribution.
The space source distribution is given by equation (20)
with $S_0=1$ and $\mu=0.3$.
The solid line is the emergent spectrum. The dashed lines represent the
contributions to the spectrum by the first nine terms of series.
\label{fig3}}

\figcaption[fig4a.eps,fig4b.eps]
{a) Plot of the emergent spectral energy flux versus photon energy
for a monochromatic injection
at  energy $E_0=10^{-4}kT_e$ and
for  three different space source distributions.
Curve 1 is for a space source distribution proportional
to the first eigenfunction.
Curves 2 and 3 are for the space source distribution given by equation (20)
with $\mu=1$ and $\mu=0.3$ respectively. The other parameters are
the same as in Figure 3. \newline
b) Same as in a), but for a blackbody input spectrum 
with $kT_{bb}=0.833$ keV.\label{fig4}}

\figcaption[fig5.eps]
{Plot of the first four eigenvalues as functions of
the accretion rate $\dot m$ in the case of $A=0$ , $kT_e=1$ keV and
$r_b=r_s$.  The solid curves correspond to $\lambda_1^2$, $\lambda_2^2$,
$\lambda_3^2$ and $\lambda_4^2$ from bottom to top.  The dashed curve gives 
the approximate analytic value of $\lambda_1^2$. \label{fig5}}

\figcaption[fig6a.eps,fig6b.eps]
{a) Plot of the energy spectral index $\alpha_1$ versus 
$\dot m$ for $r_b= r_s$ and $kT_e= 0.01$ keV (solid line), 
$kT_e= 1$ keV (dotted line) and $kT_e =3$ keV (dashed line).
\newline
b) Plot of the energy spectral index $\alpha_1$ versus
$\dot m$ for different values of the albedo $A$.
Here $kT_e=1$ keV, $r_b=r_s$ and $A$ equal to 0 (solid line),
0.5 (dotted line) and 1 (dashed line). \label{fig6}}

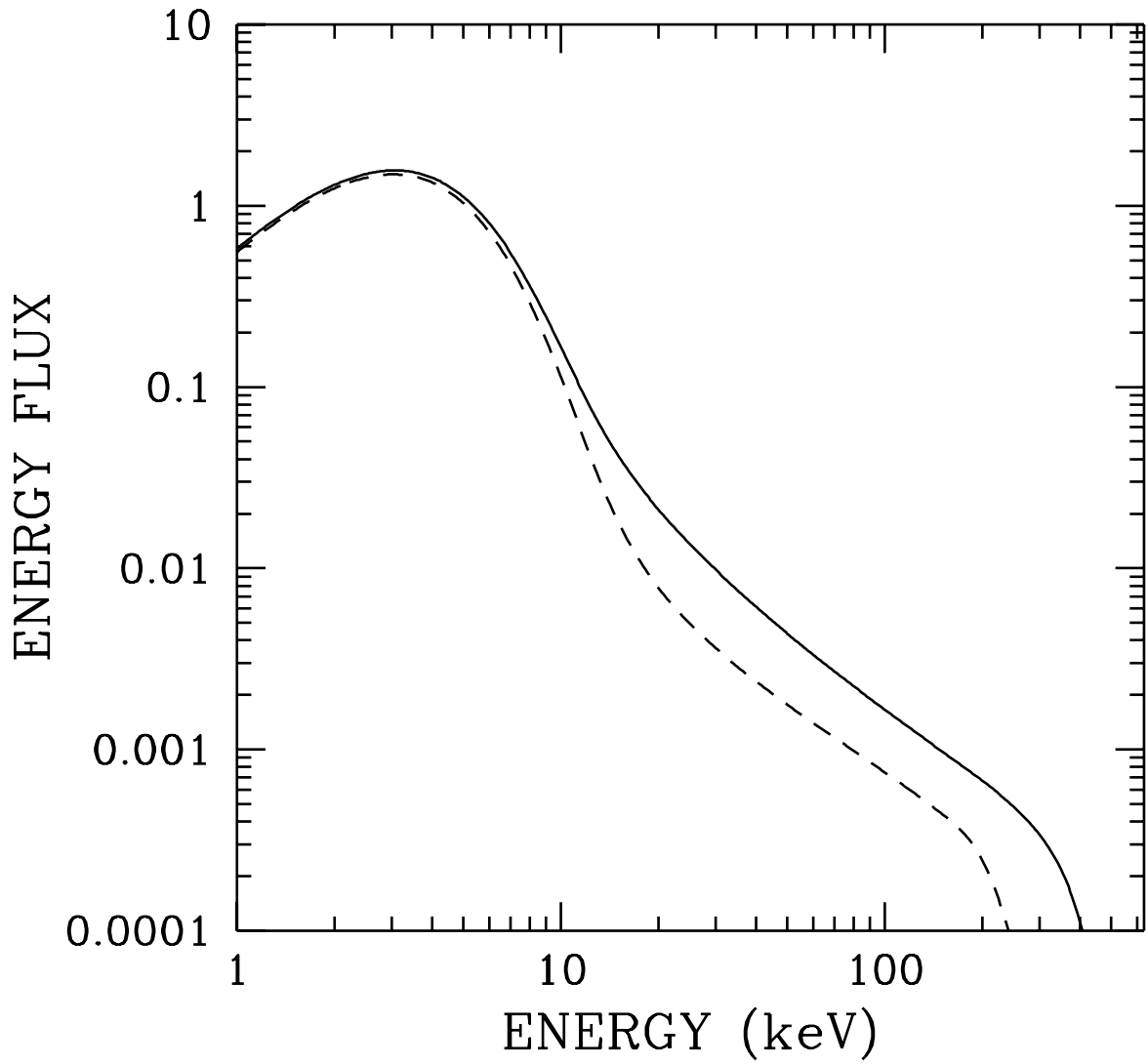
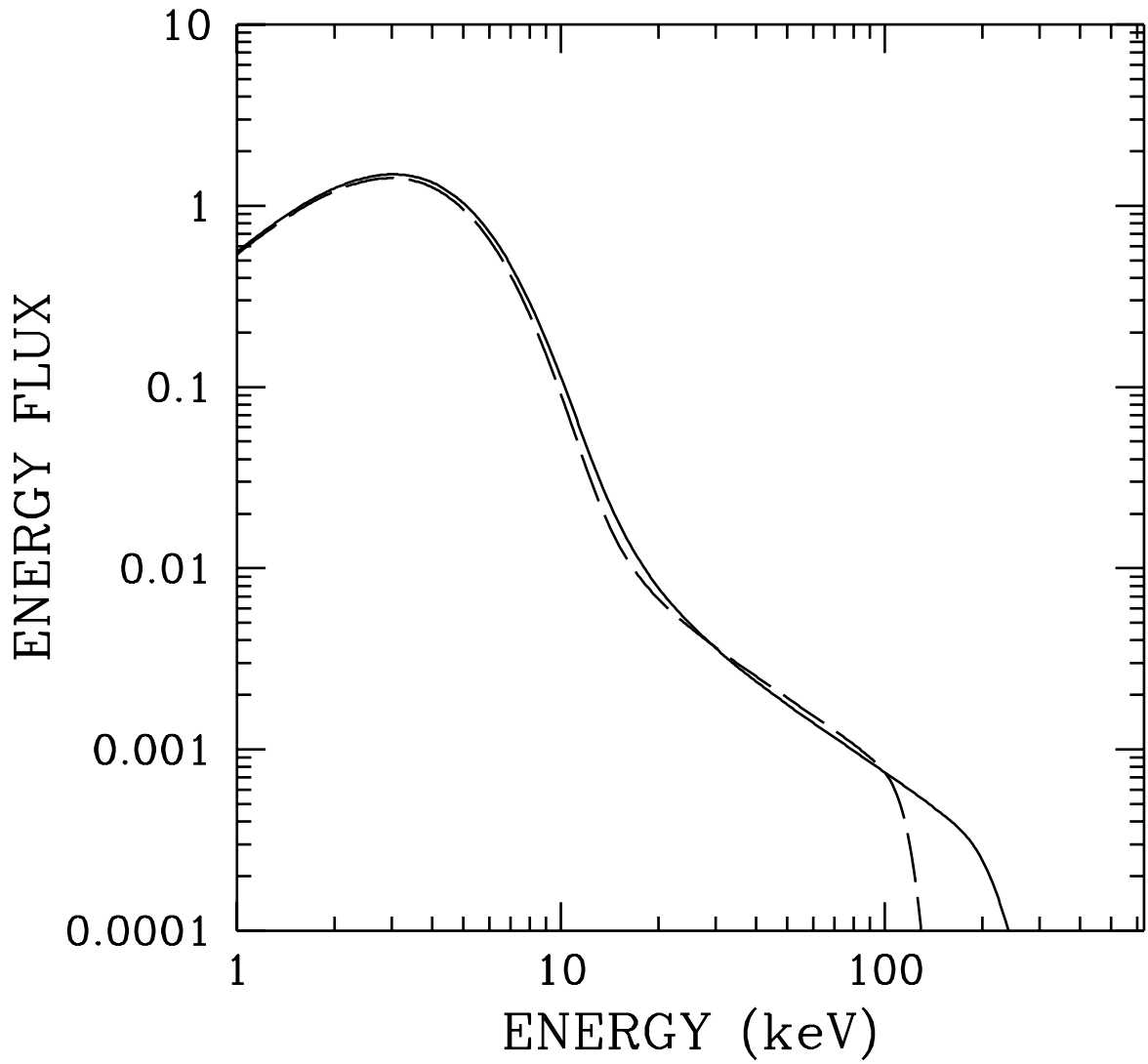
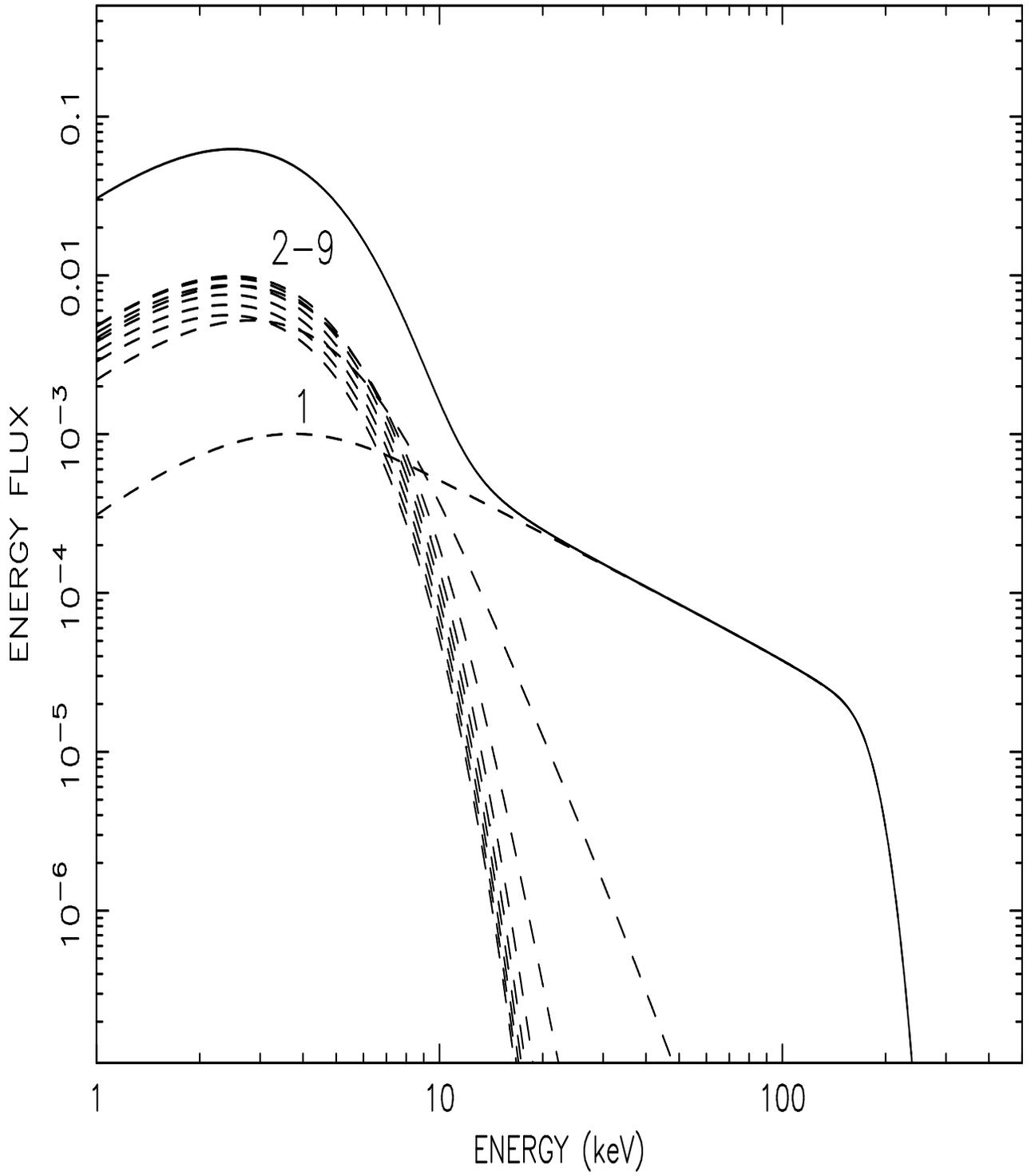
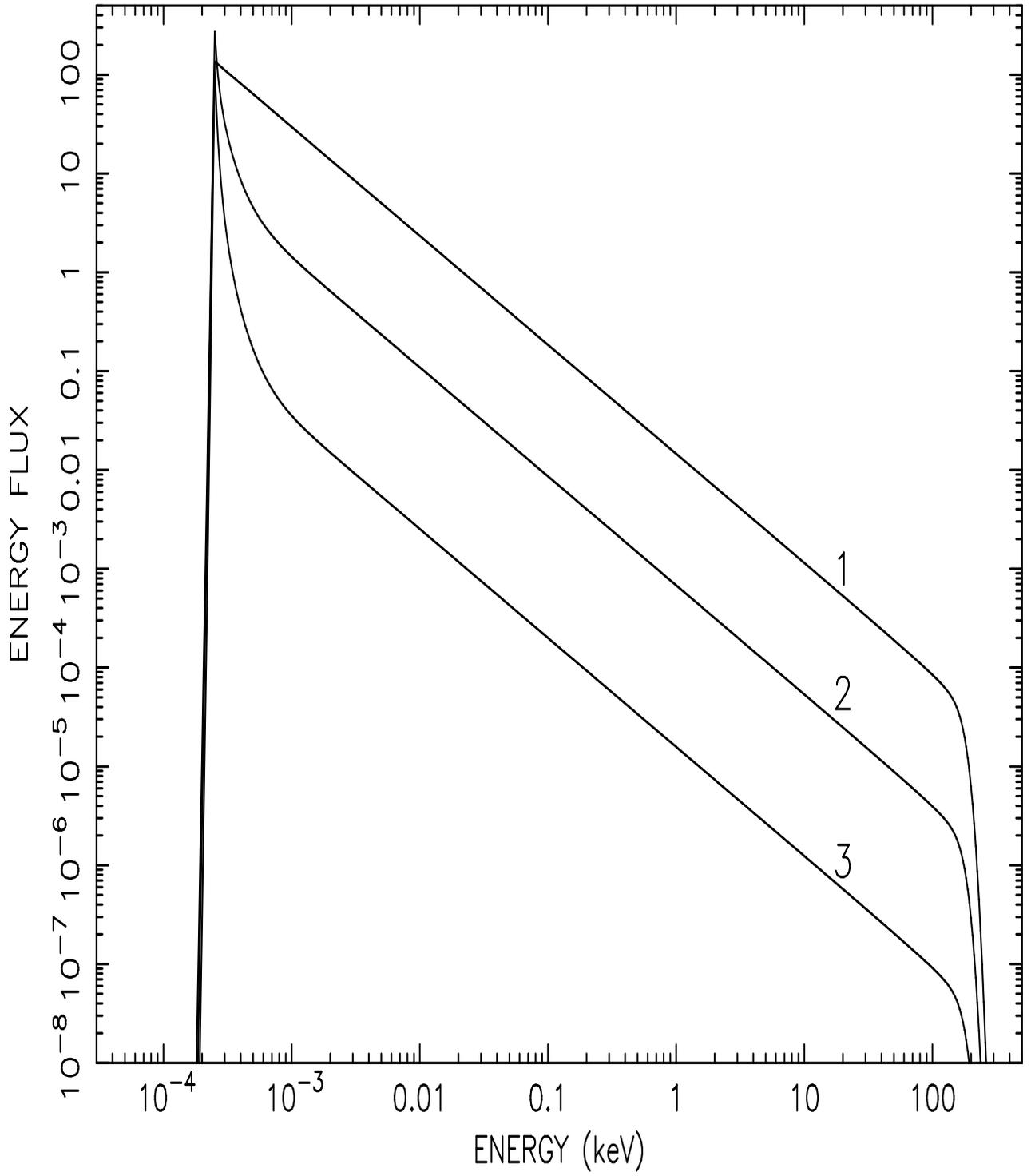
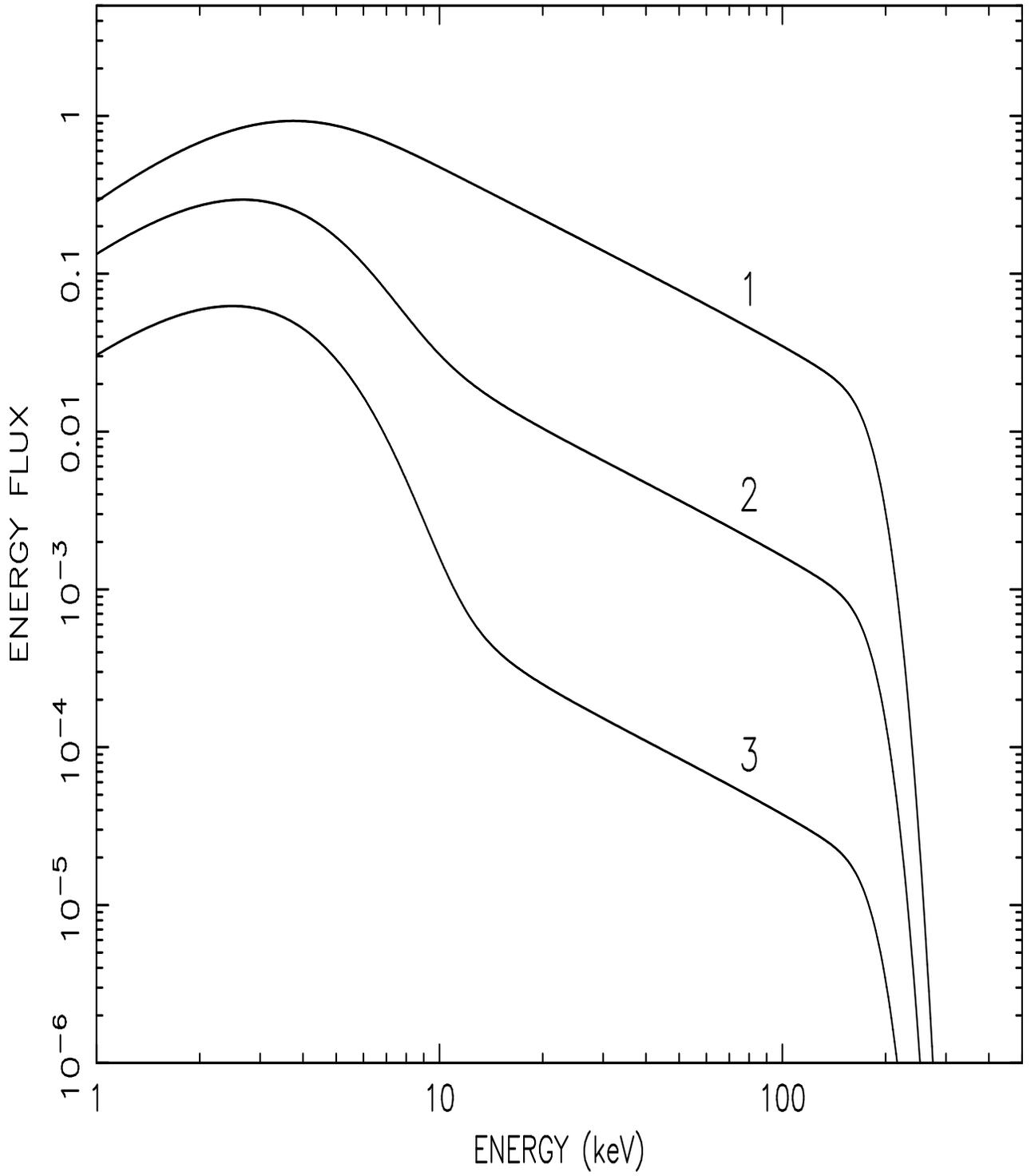
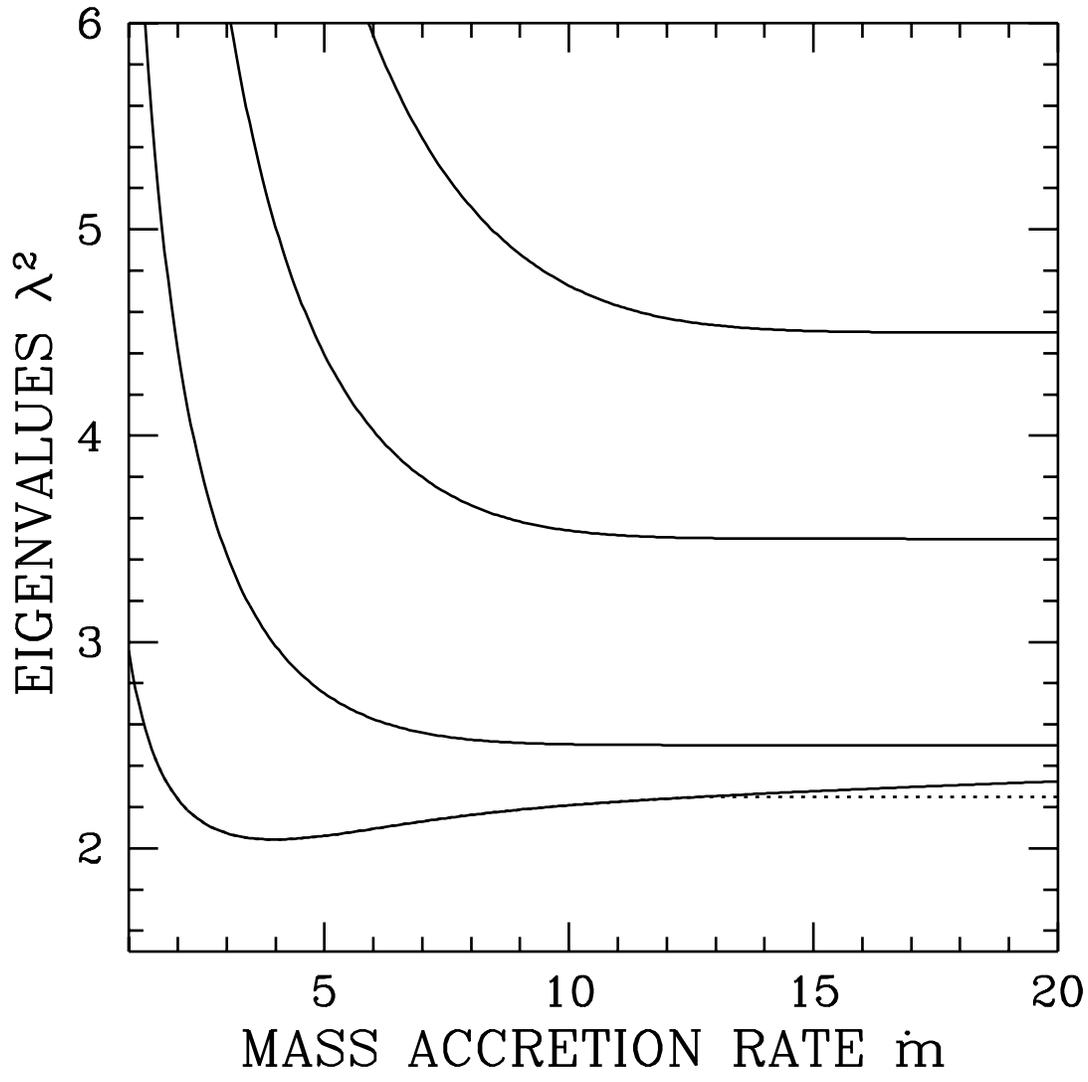
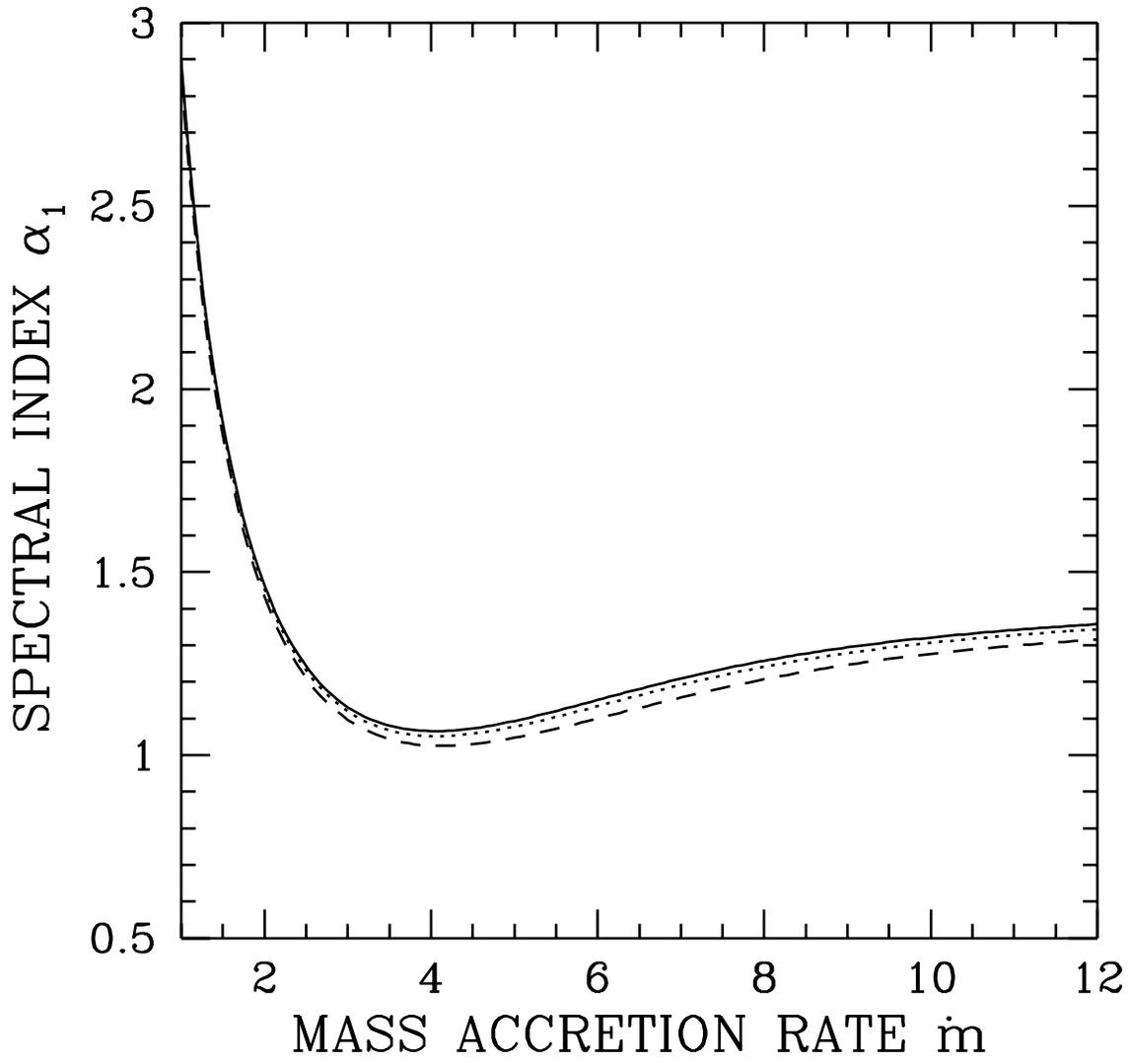
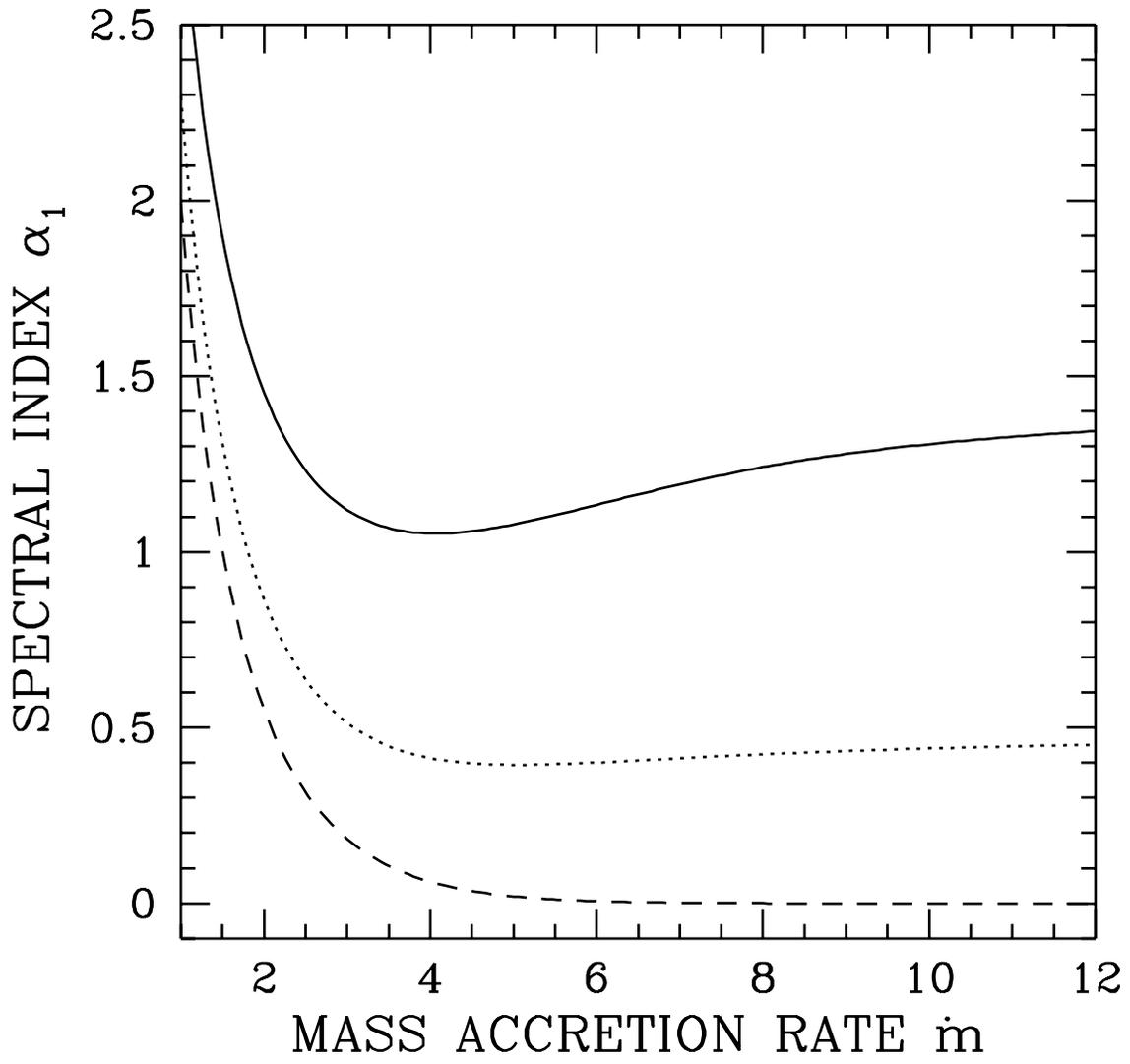
\end{document}